\documentclass[sn-mathphys,Numbered]{sn-jnl}

\usepackage{graphicx}%
\usepackage{multirow}%
\usepackage{amsmath,amssymb,amsfonts}%
\usepackage{amsthm}%
\usepackage{mathrsfs}%
\usepackage[title]{appendix}%
\usepackage{xcolor}%
\usepackage{textcomp}%
\usepackage{manyfoot}%
\usepackage{booktabs}%
\usepackage{algorithm}%
\usepackage{algorithmicx}%
\usepackage{algpseudocode}%
\usepackage{listings}%
\usepackage[utf8]{inputenc}%
\usepackage{tikz, pgfplots}%
\usetikzlibrary{positioning}%
\usetikzlibrary{backgrounds ,fit, positioning}%
\usepackage{preview}%
\usepackage{verbatim}%
\usepackage{etoolbox} 
\usepackage{listofitems} 
\usepackage{tcolorbox}
\usepackage{svg}
\usepackage{mathtools}
\usepackage{braket}
\usepackage{subcaption}

\usepackage{geometry}
\theoremstyle{thmstyleone}%
\theoremstyle{thmstyletwo}%

\theoremstyle{thmstylethree}%

\begin{document}
\title[Article Title]{Model-free Distortion Canceling and Control of Quantum Devices}

\author*[1]{\fnm{Ahmed} \sur{F. Fouad}}\email{ahmed.farouk@eng.asu.edu.eg}

\author[2]{\fnm{Akram} \sur{Youssry}}
\author[3]{\fnm{Ahmed} \sur{El-Rafei}}
\author[1]{\fnm{Sherif} \sur{Hammad}}

\affil*[1]{\orgdiv{Mechatronics Engineering Department}, \orgname{Ain Shams University}, \orgaddress{\city{Cairo}, \country{Egypt}}}

\affil[2]{\orgdiv{Quantum Photonics Laboratory and Centre for Quantum Computation and Communication Technology}, \orgname{RMIT University}, \orgaddress{\city{Melbourne}, \state{VIC 3000}, \country{Australia}}}

\affil[3]{\orgdiv{Engineering Physics and Mathematics Department}, \orgname{Ain Shams University}, \orgaddress{\city{Cairo}, \country{Egypt}}}


\abstract{
Quantum devices need precise control to achieve their full capability. In this work, we address the problem of controlling closed quantum systems, tackling two main issues. First, in practice the control signals are usually subject to unknown classical distortions that could arise from the device fabrication, material properties and/or instruments generating those signals. Second, in most cases modeling the system is very difficult or not even viable due to uncertainties in the relations between some variables and inaccessibility to some measurements inside the system. In this paper, we introduce a general model-free control approach based on deep reinforcement learning (DRL), that can work for any closed quantum system. We train a deep neural network (NN), using the REINFORCE policy gradient algorithm to control the state probability distribution of a closed quantum system as it evolves, and drive it to different target distributions. We present a novel controller architecture that comprises multiple NNs. This enables accommodating as many different target state distributions as desired, without increasing the complexity of the NN or its training process. The used DRL algorithm works whether the control problem can be modeled as a Markov decision process (MDP) or a partially observed MDP. Our method is valid whether the control signals are discrete- or continuous-valued. We verified our method through numerical simulations based on a photonic waveguide array chip. We trained a controller to generate sequences of different target output distributions of the chip with fidelity higher than 99\%, where the controller showed superior performance in canceling the classical signal distortions. 
}




\maketitle

\section{Introduction}\label{sec1}

Quantum devices promise to deliver fast computations \cite{Kielpinski_2002,doi:10.1137/S0097539796298637,Cody_Jones_2012,Shor_1997,distributed_quantum_algorithm}, precise sensing \cite{TAYLOR20161,Gross_2010,Conlon_2023,Hamley_2012,Zheng_2024,Zhuang_2024}, and secure communications \cite{Long_2007,Gisin_2007,cavaliere2020secure,paraiso2021photonic,Hu_2023,Ren_2024} compared to the current state of the art \cite{QuantumTechnology2016}. To achieve the full capability of this technology, we need to harness the functionality of these devices through proper control techniques. However, modeling and control of quantum devices is a challenging task. The fabrication process and the material properties of a quantum device could cause deviations from the intended design of the device. These imperfections introduce uncertainties in the device model and could also cause distortions to the applied control signals. Additionally, electronic and optical instruments used to generate the control signals applied to the quantum device during operation, could cause extra distortions to the control signals. This exacerbates the uncertainty in the dependence of the quantum evolution on these control signals, as in most cases the distortions model is completely unknown. These classical distortions and the uncertainty in the device model make the modeling and control procedures very challenging. In this present work, we deal with the problem of controlling closed quantum systems (specially those that are difficult to be modeled) where the control signals are subjected to classical distortions whose model may be completely unknown.

Quantum control methods can be classified as open-loop control or closed-loop control. In the open-loop approach \cite{PhysRevA.109.012617,PhysRevA.71.020302,Petruhanov_2023}, the control signals are designed beforehand and then applied to the quantum system during operation. A full accurate dynamical model of the system is mandated in this case, in order to be able to design the pulses, otherwise the pulses will not lead to the desired performance. This approach cannot accommodate for the unmodeled disturbances. On the other hand, in the closed-loop approach \cite{PhysRevA.98.052341,869285,sgroi2024reinforcement}, the control pulses are designed autonomously during operation through a feedback mechanism. This usually does not require full knowledge of the dynamical model of the system, since the feedback mechanism can compensate for it. This approach can compensate for the unexpected disturbances that may affect the system during operation, and thus it is more robust than open-loop control. A number of quantum control methods require to construct a model of the system. This model is used to predict the behavior of the system, and thus used to control it. It can also be used to compare the behaviour of the system to its design, or to understand the underlying noise process affecting it. The traditional approach to model a system is through direct physical modeling, where we look for mathematical equations that express the output signals in terms of the input and control signals. These equations will involve some unknown parameters that can be found by performing measurements on the system and using methods of parameter estimation. We call this approach the whitebox approach \cite{CompensatingNonlinearDistortions,PhysRevLett.110.040502,Complete_Characterization_Quantum_Process,PhysRevLett.114.090402,PhysRevA.64.042105}. For example in \cite{CompensatingNonlinearDistortions} the authors used the truncated Volterra series method \cite{mathews2000polynomial} to characterize non-linear distortions in controlled quantum systems. However, in many situations, the whitebox approach is not a viable option or very difficult to implement due to uncertainties in the relations between some variables, or these relations may be completely unknown. For example there may be uncertainties in the dependence of the Hamiltonian on the control signals due to the presence of unknown distortions, if any, affecting those signals. Even there could be uncertainties in the structure of the Hamiltonian itself. Additionally, there are situations where estimating the unknown parameters requires measurements that are not experimentally possible or even accessible. Moreover, the complexity of the problem increases if the physical models involve non-linear relations. The other approach that can be used for modeling and control of complex quantum systems without the need of finding exact mathematical equations, is deep supervised machine learning \cite{Goodfellow-et-al-2016}, also known as blackbox approach. Through deep supervised learning techniques, we can train neural networks (NN) to predict the output signals of the system given the input and control signals. This approach has an advantage of being capable of modeling and predicting any unknown relations between variables \cite{PhysRevX.10.011006,Papi__2022,PRXQuantum.2.010316,PhysRevA.104.052412,Ostaszewski_2019,Khait_2022,Zeng_2020}. It can even take the distortions affecting the control signals into account. However, this approach also has some drawbacks. As to reach a satisfying accuracy and to guarantee generalization of the model, a large set of labeled data is required, which is impractical in some cases. Recently, a hybrid approach, also known as graybox, has been proposed in the literature \cite{Akramchip,youssry2020characterization,perrier2020quantum,youssry2021noise,youssry2022multi,Youssry_2024,auza2024quantum}, but faces the same challenges of supervised learning approach. Namely in \cite{Akramchip}, the authors used recurrent neural networks to model and control a photonic waveguide array chip, but the model was trained to predict the output for control signals of square waveform shape only.

Alternatively, there are control methods that aim directly to control the system without first modeling it. For example, dynamical decoupling and dynamically-corrected gates \cite{PhysRev.94.630,Viola_1999,Biercuk_2011,Khodjasteh_2009}, as well as direct gradient-based optimization, such as the GRAPE algorithm \cite{Khaneja_2005} and its variants \cite{de_Fouquieres_2011, Ciaramella_2015,Abdelhafez_2019,Leung_2017} work on optimizing the fidelity to some target with respect to control. Only the dependence of the Hamiltonian on the control should be known in this case. Even in situations where this dependence is unknown, for instance if the control signals are subjected to unmodeled classical distortions, the fidelity and/or its gradient can be computed iteratively from experimental data. After each iteration the control signals are optimized and directly applied to the physical system for the next iteration, where the physical system becomes part of a feedback architecture for designing the pulses without a need for a model. This approach is sometimes referred to as ``learning quantum control" \cite{Wu_2019,Li_2017,Chen_2020,Yang_2020}. Reinforcement learning (RL) methods are also employed in quantum control. They are model-free and are also considered as a learning quantum control approach. RL becomes yet more powerful when combined with deep neural networks, which is known as deep reinforcement learning (DRL) \cite{ModernDrlAlgos,SHAKYA2023120495,9904958,8103164}. DRL techniques enable intelligent decision-making in complex environments. They can train an agent (controller) to learn an optimal control policy through trial and error, similar to how humans learn from experience, by interacting with its environment (system) in the form of a black-box. It observes the environment current state and takes actions based on this state. After each action, the agent receives feedback in the form of rewards or penalties. The objective of the agent is to learn a policy that maximizes the cumulative reward, which represents the target problem, over time. The training of the agent does not require any labeled data, as the data used for training is automatically generated by the agent during training through sampling from the environment.

DRL has been employed in the past few years in quantum systems and technology field for quantum error correction \cite{PhysRevX.8.031084,Nautrup2019optimizingquantum}, quantum state transfer \cite{Porotti_2019,PhysRevA.103.L040401,PAPARELLE2020126266}, quantum metrology \cite{PhysRevA.103.042615, PhysRevLett.131.073201}, quantum state preparation and engineering \cite{Zhang_2019,Mackeprang_2020,Haug_2021,Porotti2022deepreinforcement,PhysRevResearch.2.033295,Zen:2024vgp,Xu:2023rvi,PhysRevLett.126.060401,Alam_2023}, and quantum control \cite{PhysRevLett.125.100401,Giannelli_2022,An_2019,Guatto2024ImprovingRO,Zhou2023AuxiliaryTD,Niu_2019,PhysRevApplied.18.024033,Borah_2021,Brown2021ReinforcementLP,August_2018,pmlr-v107-yao20a,Wauters_2020,Lockwood2021OptimizingQV}. Focusing on the implementation of DRL in quantum control and quantum state preparation, we found the following gaps in the current literature. Particularly, some of the existing work
\begin{enumerate}
	\item do not take into account the classical distortions, mentioned earlier, that could affect the control signals, which renders this work experimentally impractical \cite{PhysRevLett.125.100401,Mackeprang_2020,Haug_2021,Guatto2024ImprovingRO,Porotti2022deepreinforcement,An_2019,Giannelli_2022,Borah_2021,Zhou2023AuxiliaryTD}.
	\item focus on driving the system evolution to a single fixed target state, which makes them not general enough and of limited usage \cite{August_2018,Brown2021ReinforcementLP,Xu:2023rvi,pmlr-v107-yao20a,Borah_2021,Mackeprang_2020,Guatto2024ImprovingRO,Porotti2022deepreinforcement,An_2019}.
	\item deal only with quantum control problems that can be modeled as a Markov decision process, which is usually not the situation (as this need full observability of the system state) \cite{Alam_2023,Giannelli_2022}.
	\item use discrete action space for the control problem, which is not suitable for many applications that use continuous-valued control signals \cite{Mackeprang_2020,Alam_2023}.
\end{enumerate} 

In this paper, we aim to close those gaps in the literature by proposing a general control approach based on DRL. This approach works for any closed-quantum system, taking into account the classical distortions that could affect the control signals. Through our model-free universal approach, we control the state probability distribution of the system, and drive it to different target distributions. We are using closed-loop control, as we continuously monitor the evolution of the state probability distribution by direct measurement. Our controller is a deep NN trained using REINFORCE policy gradient algorithm \cite{Williams_1992,ModernDrlAlgos}. This algorithm works whether the control problem can be modeled as an MDP or not (i.e., partially observed Markov decision process (POMDP)). In this work, we are employing a novel controller architecture which, to the best of our knowledge, was not employed before in the literature. The proposed architecture comprises multiple NNs. This enables accommodating as many different target state distributions as desired, without increasing the complexity of the network or its training process. Our method is valid for both discrete and continuous action.


We will verify our method through numerical simulations based on the device introduced in \cite{Akramchip}. This device is a voltage-controlled optical waveguide array chip, where a laser beam is injected into the input of one of the array waveguides, and only the output optical power distribution across all the waveguides can be measured. The material properties of this chip cause distortions to the applied control voltages. We will show the results of implementing a controller to control the probability distribution of the output state of the chip, while compensating for these distortions. This controller can drive the chip output to different target probability distributions.

The structure of the remainder of the paper is as follows. In Section \ref{sec2}, we mathematically formulate the problem we are trying to solve. Next in Section \ref{sec3}, we present our method, where our novel controller architecture is introduced in Section \ref{subsec1}. After that, we present the numerical simulation results of applying our method to the aforementioned chip in Section \ref{sec4}. Then, we discuss the significance of these results and some of the advantages of our method in Section \ref{sec5}. Finally, we conclude our paper in Section \ref{sec6}.

\section{Problem Statement}\label{sec2}
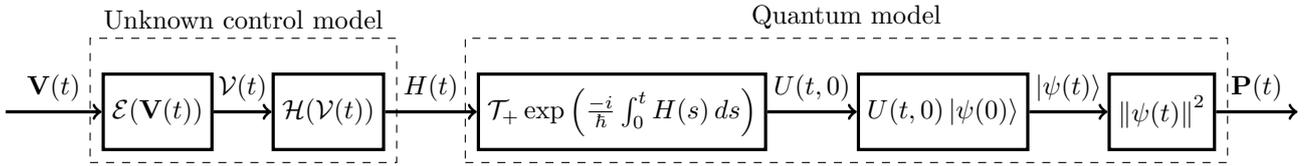
\begin{figure}
	\centering
\begin{tikzpicture}[
	SIR/.style={rectangle, draw=black, very thick, minimum size=10mm},
	outer/.style={rectangle, draw=black, dashed, inner xsep=1ex, inner ysep=2ex, yshift=1ex,
		fit=#1}
	]
	\node[SIR]    (Distrotions)                              {$\mathcal{E}(\textbf{V}(t))$};
	\node[SIR]    (HamDependence)       [right=of Distrotions] at (0.5,0){$\mathcal{H}(\mathcal{V}(t))$};
	\node[SIR]    (UnitDependence)       [right=of HamDependence]at (3.2,0) {$\mathcal{T}_{+}\exp{\left(\frac{-i }{\hbar}\int_{0}^{t}H(s)\,ds\right)}$};
	\node[SIR]    (timeevolution)       [right=of UnitDependence] at (8.2,0){$U(t,0)\ket{\psi(0)}$};
	\node[SIR]    (measurement)       [right=of timeevolution] at (11.5,0) {$\left\|\psi(t)\right\|^2$};
	
	\scoped[on background layer]
	\node [outer=(Distrotions) (HamDependence),
	label={Unknown control model}] {};
	
	\scoped[on background layer]
	\node [outer=(UnitDependence) (timeevolution) (measurement),
	label={Quantum model}] {};
	
	\draw[->,very thick] (-2,0) to node[above] {$\textbf{V}(t)$} (Distrotions.west);
	\draw[->, very thick] (Distrotions.east)  to node[above]  {$\mathcal{V}(t)$} (HamDependence.west);
	\draw[->, very thick] (HamDependence.east)  to node[above] {$H(t)$} (UnitDependence.west);
	\draw[->, very thick] (UnitDependence.east)  to node[above] {$U(t,0)$} (timeevolution.west);
	\draw[->, very thick] (timeevolution.east)  to node[above] {$\ket{\psi(t)}$} (measurement.west);
	\draw[->, very thick] (measurement.east)  to node[above] {$\textbf{P}(t)$} (15,0);

\end{tikzpicture}
    \vspace{0.1cm}
	\caption{The Block diagram shows the relation between the control signals $\textbf{V}(t)$ and the measured probability distribution $\textbf{P}(t)$ of the quantum state of the system. The unknown classical distortions $\mathcal{E}$ change the control signals $\textbf{V}(t)$ into distorted control signals $\mathcal{V}(t)$ before affecting the system Hamiltonian $H(t)$. The dependence $\mathcal{H}$ of $H(t)$ on $\mathcal{V}(t)$ is also unknown. The evolution unitary operator $U(t,0)$, which is the time-ordered matrix exponential of the system Hamiltonian $H(t)$, acts on the quantum system state to evolve it from $\ket{\psi(0)}$ to $\ket{\psi(t)}$. We obtain $\textbf{P}(t)$ by applying measurement to $\ket{\psi(t)}$.
}\label{fig1}
\end{figure}

The objective of this work is to control the evolution of closed quantum systems (specially systems that are difficult to be modeled), where the control signals that drive the system Hamiltonian are subjected to unmodeled classical distortions. The dependence of the Hamiltonian on the control signals, even if there were no distortions at all, could be nonlinear or even unknown. The challenges to be tackled in this paper are as follows.
\begin{enumerate} [1.]
	\item Firstly, the classical distortions that affect the control signals applied to the quantum system to drive its evolution. In most cases these distortions are very difficult to be modeled. These distortions could arise from the device fabrication \cite{Yamada_1981,LiNbO3opticalintensitymodulators8}, material properties \cite{Akramchip,Yamada_1981,LiNbO3opticalintensitymodulators8} and/or the device operation including the external electronic and optical instruments generating the control signals \cite{CompensatingNonlinearDistortions,PhysRevLett.110.040502,PhysRevB.80.220506,Nakamura_1999,2005cond.mat..8587Y}. These distortions distort the control signals before they affect the Hamiltonian. Thus, the waveform and the shape of the actual signals affecting the Hamiltonian are different from those of the ones being applied to the system. 
	\item Secondly, the difficulty of identifying the system or modeling the uncertainties regarding the structure of the Hamiltonian and its dependence on the control signals. This difficulty could arise from the inaccessibility to some measurements inside the system \cite{Akramchip,Youssry_2024}. In most cases, only the probability distribution of the system state can be observed. Therefore, characterizing some parameters that determine, for example, the dependence of the Hamiltonian on the control signals or the distortions model becomes impossible.
\end{enumerate}
The evolution of the state $\ket{\psi(t)}$ of a closed quantum system at time $t$ from an initial state $\ket{\psi(0)}$ is given by \begin{equation}
	\ket{\psi(t)} = U(t,0)\ket{\psi(0)}\label{eq1}.
\end{equation}
The evolution unitary operator $U(t,0)$ is a function of the system Hamiltonian $H(t)$ as described by \begin{equation}
	U(t,0) = \mathcal{T}_{+}\exp{\left(\frac{-i }{\hbar}\int_{0}^{t}H(s)\,ds\right)}\label{eq2},
\end{equation} where $\mathcal{T}_{+}$ is the time-ordering operator.
In this paper, we are trying to control the state probability distribution $\textbf{P}(t)$ of the quantum system through applying external control signals $\textbf{V}(t)$ and monitoring the state probability distribution $\textbf{P}(t)$ which is a measurable quantity. The relationship between $\textbf{P}(t)$ and $H(t)$  is inherently non-linear. Moreover, the classical distortions $\mathcal{E}$ change $\textbf{V}(t)$ into distorted control signals $\mathcal{V}(t)$ before affecting $H(t)$, even the dependence $\mathcal{H}$ of $H(t)$ on $\mathcal{V}(t)$ is unknown. The relation between $\textbf{V}(t)$ and $\textbf{P}(t)$ can be summarized in the block diagram shown in Figure \ref{fig1}. It is obvious that our control problem is highly non-linear and complex to be solved using classical control methods, and thus, machine learning techniques would be a proper approach.

All the above issues are addressed in our proposed method that will be introduced shortly in the next section. In our approach we use a policy gradient DRL algorithm to obtain a controller (policy) for the quantum system. This algorithm is model-free. It deals with the system as a black box.

\section{Methods}\label{sec3}
To tackle the challenges mentioned earlier, a model-free control approach is proposed. This approach employs a closed-loop control scheme through utilizing a feedback to continuously monitor the system state probability distribution $\textbf{P}(t)$ as it evolves. The controller is an NN that will be trained using REINFORCE policy gradient DRL algorithm \cite{Williams_1992,ModernDrlAlgos} through direct interaction with the system to be controlled.


\subsection{Controller Architecture}\label{subsec1}
\begin{figure}
	\centering
	\includegraphics[scale=0.7]{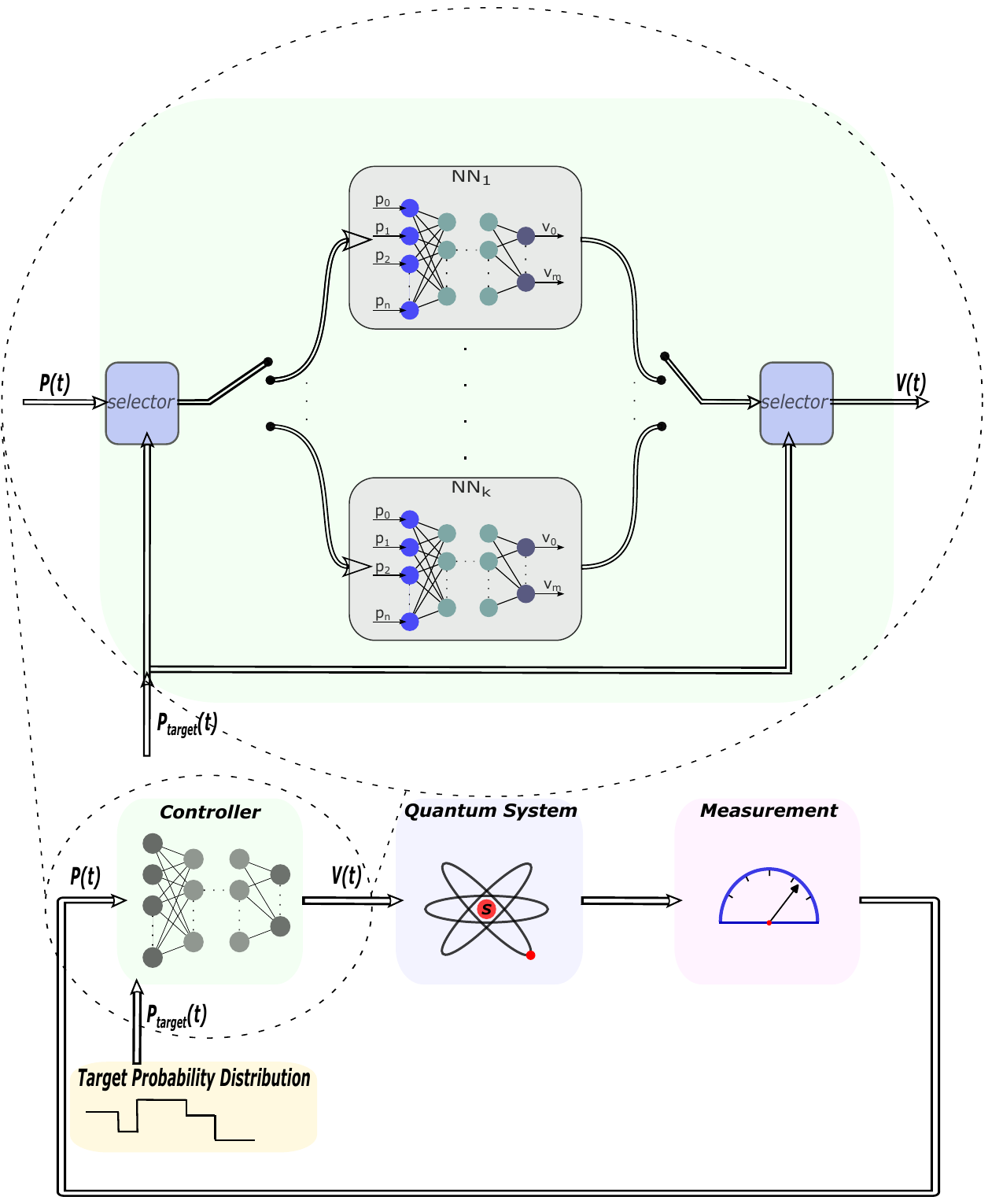}
	\caption{The control loop used to control the quantum system. The inset shows the controller architecture. The controller (represented by the set of NNs) outputs the control signals $\textbf{V}(t)$ that are applied to the system. The measured state probability distribution of the system $\textbf{P}(t)$ is looped back to be the input to the controller along with the target distribution $\textbf{P}_{\text{target}}(t)$ desired at the moment. The controller consists of a set of NNs. It comprises a fully-connected feedforward NN for each desired target state probability distribution that we want to achieve. This NN can bring the system from an initial state probability distribution to the corresponding target distribution. The controller has a selector that selects the corresponding NN according to the target state probability distribution desired at the moment. This proposed controller architecture can handle any number of desired target distributions.}\label{fig2}
\end{figure}

Training an NN to bring the quantum system from an initial state probability distribution to all possible target distributions, is a very complex task that will increase the complexity and the size of the NN and make the training process very difficult. We adopt another approach, where the controller is not just one NN, but it consists of a set of NNs as shown in Figure \ref{fig2}. The controller comprises a separate fully-connected feedforward NN for each desired target state probability distribution. This NN can bring the system from an initial state probability distribution to the corresponding target distribution. The controller has a selector that selects the corresponding NN according to the target state probability distribution desired at the moment. This proposed controller architecture can handle any number of desired target distributions. Practically speaking, it is not needed to drive a quantum system to all possible state probability distributions, but only to a finite set of target distributions depending on the application. For example, if we control a device to act as a configurable quantum gate, we do not have to achieve all possible gates, they are infinite, but we only need to achieve a set of desired target gates. Namely, if we have $k$ desired target state probability distributions (gates), then our controller will consist of $k$~NNs, and if we want to drive that system into a sequence of these distributions (gates), the control signals will be computed by the selected NN that corresponds to the desired target distribution (gate) at the moment. Our controller could be thought of as $k$ different controllers each dedicated to achieve a certain target distribution. Our controller can generalize to any number of target distributions.

The setup shown in Figure \ref{fig2} shows the proposed control loop. This is also the same setup used to train the controller. The setup is as follows.
\begin{enumerate} [1.]
	\item The controller (represented by the set of NNs) outputs the control signals $\textbf{V}(t)$ that are applied to the system.
	\item The state probability distribution of the system $\textbf{P}(t)$ will be looped back to be the input to the controller along with the target distribution $\textbf{P}_{\text{target}}(t)$ desired at the moment.
\end{enumerate}
\subsection{Algorithm Design}\label{subsec2}

From DRL perspective our control problem can be formulated as follows.
\begin{enumerate} [1.]
	\item The controller represents the agent with policy $\pi_{\theta}(\textbf{a}_{t}|\textbf{s}_{t})$ (which is represented by the set of NNs).
	\item The quantum system represents the DRL environment $p(\textbf{s}_{t+1}|\textbf{a}_{t},\textbf{s}_{t})$ where the DRL environment state $\textbf{s}$ is represented by the quantum state probability distribution $\textbf{P}(t)$, and the action $\textbf{a}$ taken by the agent is represented by the control signals $\textbf{V}(t)$ generated by the controller.
\end{enumerate}
The agent control task to reach the target probability distribution is divided into a number $T$ of time steps. In DRL context, these steps collectively is called an episode. The agent takes an action at each time step based on the environment observed state, which induces a transition of the environment to a new state (the state of the next step).

If we take a closer look, we will find that our control problem in this formulation is not an MDP. However, most of the DRL algorithms work best for MDPs, where the DRL environment next state $\textbf{s}_{t+1}$ depends only on the current state $\textbf{s}_{t}$ and the current action $\textbf{a}_{t}$ regardless of the history of the state-action pairs. In our particular control problem, we cannot claim that the next probability distribution of the quantum state(next DRL state) depends only on the current probability distribution of the quantum state (current DRL state) and the current applied control signals (current action), due to the presence of classical distortions $\mathcal{E}$. These distortions affect the applied control signals before they actually drive the system Hamiltonian, where the quantum state evolution, and thus the state probability distribution, depends on the system Hamiltonian. These distortions could be linear or non-linear, and even in the linear case, it will be modeled as a linear-time-invariant (LTI) system whose input is the applied control signals $\textbf{V}(t)$, and the response of this LTI system $\mathcal{V}(t)$ is the one actually driving the Hamiltonian. An LTI system has memory, which means that its response depends on the history of the input not just the current value of the input, and thus the system Hamiltonian will depend on the history of actions (applied control signals) not just the current action. Consequently, the next state probability distribution will not depend only on the current state-action pair only, but it will depend on the history of actions (history of the control signals). Therefore, our control problem is not an MDP but it is a POMDP. That is why we use the REINFORCE algorithm, as it does not require the process to be MDP (works for POMDP as well as MDP), which is known from the derivation of the gradient estimation equation of this algorithm \cite{UC_Berkly_notes}.



\subsection{Controller Training}\label{subsec3}
In DRL, the NN (policy) learning is guided by the reward function which rewards/penalizes the NN at each time step $t$ if it takes the right/wrong action for the input current state. The reward function is crucial to have a successful training in DRL. To train our controller, each NN in the controller is trained separately using a reward function $r(\textbf{a}_{t},\textbf{s}_{t})$. The training of each NN goes as follows. A number $N$ of episodes (trajectories $\tau^i$) are run in the system (i.e., unrolling the policy in the DRL environment), then the NN is updated based on the reward achieved in these episodes by taking a policy gradient ascent step using the REINFORCE algorithm gradient estimation formula shown in Equation \ref{eq3} \cite{UC_Berkly_notes,ModernDrlAlgos}. At the beginning of each episode, the system (DRL environment) is reset to a certain initial state $\textbf{s}_{0}$. During the episode, the NN (policy) takes the current probability distribution of the quantum state of the system ($\textbf{s}_{t}$) as input and outputs the control signals ($\textbf{a}_{t}$) which is applied to system. Then the evolved probability distribution ($\textbf{s}_{t+1}$) due to this action is taken as the input of the next step of the episode and so on until the episode is over. The reward function $r(\textbf{a}_{t},\textbf{s}_{t})$ at each step $t$ is calculated based on the absolute difference between the corresponding target quantum state probability distribution $\textbf{P}_{\text{target}}$ and the quantum state probability distribution $\textbf{s}_{t+1}$ evolved due to applying action $\textbf{a}_{t}$. The learning process continues this way for a number of updates until the NN (policy) reaches the desired accuracy. This way the NN learns to drive the system step by step during the episode time to reach the target probability distribution and cancel the signal distortions by selecting the proper control signals at each step. One of the advantages of this training scheme is that the training samples are automatically generated by the NN (agent) through sampling from the actual environment. We do not have to collect or design the training data set prior to training, to guarantee generalization as in deep learning schemes. Once the training is finished successfully, the trained NN has learned an efficient policy $\pi_{\theta}(\textbf{a}_{t}|\textbf{s}_{t})$ which selects the best action $\textbf{a}_{t}$ at the current state $\textbf{s}_{t}$ to drive the system to the corresponding target quantum state probability distribution $\textbf{P}_{\text{target}}$ (on which the NN is trained to achieve) before the episode time limit, even if the episode starts at a DRL state different from the reset state $\textbf{s}_{0}$ used in training. This obtained policy will be able to generalize to states unseen during training, like starting the episode from a different initial state. This training scheme is summarized in Algorithm \ref{alg:1}.

However, the REINFORCE algorithm suffers from a relatively high variance in the gradient estimates used for updating the policy \cite{ModernDrlAlgos}. To overcome this issue, we used some known techniques. Namely, we applied the reward-to-go technique by using the sum of upcoming rewards at each step of the episode and ignoring past rewards \cite{ModernDrlAlgos,UC_Berkly_notes}. In addition, we used a discount factor $\gamma$ to make the agent focus more on the rewards that are closer in time than those that are further in the future, which is also known to reduce variance \cite{ModernDrlAlgos,UC_Berkly_notes}. Another known technique that we used to reduce variance, is subtracting a baseline function $b(\textbf{s})$ from the total reward of each generated trajectory during the training process \cite{ModernDrlAlgos,UC_Berkly_notes}. This baseline function must be independent of the action $\textbf{a}$ but could depend on the state $\textbf{s}$. This could be done by subtracting a constant from the total reward of each trajectory, so that the good trajectories would have positive rewards and bad trajectories would have negative reward, which makes it easier to update the policy to increase the likelihood of good trajectories and decrease the likelihood of bad ones.
\section{Results}\label{sec4}

In this section, we validate our method on the quantum system presented in \cite{Akramchip}. In \cite{Akramchip}, the authors introduced a voltage-controlled integrated optical waveguide array chip with a reconfigurable Hamiltonian. A laser beam is injected into the input of one of the array waveguides, and only the output optical power distribution across all the waveguides can be measured. 
\begin{algorithm}
	\caption{REINFORCE Algorithm}\label{alg:1}
	\begin{algorithmic}
		\While {NN accuracy $\leq$ desired accuracy}
		\State $i \gets N$
		\While{$i \neq 0$}
		\State Reset the DRL environment (the quantum system) to a definite initial
		\State state $\textbf{s}_{0}$.
		\State Sample $\tau^i$ from $\pi_{\theta}(\textbf{a}_{t}|\textbf{s}_{t})$ (run $\pi_{\theta}$ in the DRL environment).
		\State $i \gets i - 1$
		\EndWhile
		\State Calculate gradient: \begin{equation} \nabla_{\theta}J(\theta) \approx \frac{1}{N} \sum_{i=1}^{N}(\sum_{t=0}^{T-1} \nabla_{\theta}\log\pi_{\theta}(\textbf{a}_{t}^i|\textbf{s}_{t}^i) ((\sum_{\tilde{t}= t}^{T-1}\gamma^{\tilde{t}-t}r(\textbf{a}_{\tilde{t}}^i,\textbf{s}_{\tilde{t}}^i))-b(\textbf{s}_{t}^i))). \label{eq3} \end{equation}
		\State Take gradient ascent step: $\theta \gets \theta + \eta\nabla_{\theta} J(\theta)$, $\eta$ is the learning rate.
		\EndWhile
	\end{algorithmic}
\end{algorithm}
A chip with two waveguides is described quantum mechanically with the computational basis encoding the presence of photons in each waveguide where the state $\ket{0} = [1,0]^T$ encodes a photon present at the first waveguide and, the state $\ket{1} = [0,1]^T$ encodes a photon in the second waveguide. The light power distribution at the inputs of the chip waveguides represents the initial quantum state $\ket{\psi(0)}$ of the system, while the light power distribution at the outputs of the chip waveguides represents the final quantum state $\ket{\psi(t_{l})}$ of the system, where $t_{l}$ is the time taken by the light to cross the chip of length $l$. The behavior of the chip when light propagates along the waveguides represents the evolution of the system from $\ket{\psi(0)}$ to $\ket{\psi(t_{l})}$. There are two electrodes through which we change the external applied voltage $\textbf{V}(t)$ across the first and second waveguides respectively. These applied voltage will suffer from classical distortions introduced by the material properties of the chip. The Hamiltonian $H$ of the chip is a function of the distorted voltages $\mathcal{V}(t)$. These distortions cannot be modeled or measured in any way. Even the structure of the Hamiltonian and the exact relation between it and the distorted voltages is unknown, since we do not have access to measurements inside the chip. Thus identifying the chip as a white box is almost impossible. The time ordered evolution unitary operator given in Equation \ref{eq2}, will reduce in this case to 
 \begin{equation}
	U = \exp{(-iHt_{l})}\label{eq4}.
\end{equation}   
Since the time scale of changing the voltage $\textbf{V}(t)$ is much slower than the time scale of the photon travel across the chip, each photon can see only one time-independent Hamiltonian from the moment it enters the chip until the moment it reaches the output. This allows us to write the evolution as the matrix exponential of the Hamiltonian as in Equation \ref{eq4}. The voltage applied to each electrode should not exceed an absolute value of 10~V otherwise the chip could be damaged \cite{Akramchip}.

In the rest of this section, we show the results of applying our method to a two-waveguide chip where we take the measured output power distribution of the chip $[\alpha,\beta]^T$ as the DRL environment state $\textbf{s}$, and the external contol voltages $\textbf{V}$ as the action $\textbf{a}$. We applied our method to the simulator created for the chip by the authors in \cite{Akramchip}, using the same parameters the authors used while applying their method to this simulator. This simulator generates the waveguide power distribution given a set of control voltages, where the classical distortions that distort these control voltages are modeled as an LTI system with a second-order transfer function. The output light power distribution of the chip $[\alpha,\beta]^T$ is assumed to be normalized.

For implementation we considered five target output power distributions ([0,1], [0.2,0.8], [0.5,0.5], [0.8,0.2], and [1,0]) on which we will train our controller to achieve, where the light distribution at the chip inputs is fixed to [0,1]. Thus our controller consists of five NNs. These five distributions spans the whole spectrum of the output power distribution of the chip from [0,1] to [1,0]. Each NN in the controller consists of 4 hidden layers, each with 128 nodes with hyperbolic tangent activation function. The size of input to the NN is 50 which is the output power distribution of the first waveguide of the chip for the current DRL step sampled over 50 points. We only considered the output of the first waveguide since the output power distribution is already normalized. Consequently, the output power of the second waveguide will not give new information. The NN outputs four parameters which are the mean and variance of two gaussian distributions representing the two control voltages $\textbf{V}(t)$. The values of the mean are scaled between -10~V and~10 V using hyperbolic tangent activation function.

\subsection{Training and Evaluation}\label{subsec4}
As mentioned in the methods section, each target distribution dedicated NN is trained separately. For training an NN to achieve a target power distribution $[\alpha_{\text{target}},\beta_{\text{target}}]^T$, we used a reward function:
\begin{equation}
	r(\textbf{a}_{t},\textbf{s}_{t}) = -\,c_{\text{target}}\,(|\alpha_{t+1}^{*} - \alpha_{\text{target}}| - m_{\text{target}}) \label{eq5}, 
\end{equation} 
where $[\alpha_{t+1}^{*},\beta_{t+1}^{*}]^T$ is last sample of the chip output power distribution generated by applying the action $\textbf{a}_{t}$ at the current DRL step (since we sample the chip output response during each time step over 50 points). $m_{\text{target}}$ is a constant value subtracted from the absolute difference $|\alpha_{t+1}^{*} - \alpha_{\text{target}}|$ to center the reward range around zero, so that we could have positive and negative rewards. This enhances the process of policy training, as it makes it easier to update the policy to increase the likelihood of good trajectories with positive rewards and decrease the likelihood of bad trajectories with negative rewards. For example if $\alpha_{\text{target}} = 0.8$, the absolute difference ranges from 0 to 0.8, then $m_{\text{target}}$ should equal 0.4, so that the range will become from -0.4 to 0.4 (the reward range will become from $-0.4c_{\text{target}}$ to $0.4c_{\text{target}}$). This centering technique we just mentioned is equivalent to using a reward function $r(\textbf{a}_{t},\textbf{s}_{t}) = -\,c_{\text{target}}\,|\alpha_{t+1}^{*} - \alpha_{\text{target}}|$, with baseline function $b(\textbf{s}) = -\,c_{\text{target}}\,m_{\text{target}}\,\sum_{\tilde{t}= t}^{T-1} \gamma^{\tilde{t}-t} $. Since for different target distributions, we have different value ranges for the quantity $-\,(|\alpha_{t+1}^{*} - \alpha_{\text{target}}| - m_{\text{target}})$, we use a constant value $c_{\text{target}}$ to scale the range of rewards to be from -25 to 25 to standardize the reward range between different targets. This reward range turned out to be the best performing based on our experiments.

\begin{figure}
	\centering
	\begin{subfigure}{1\linewidth}
		\centering{
        \includegraphics[scale =0.348]{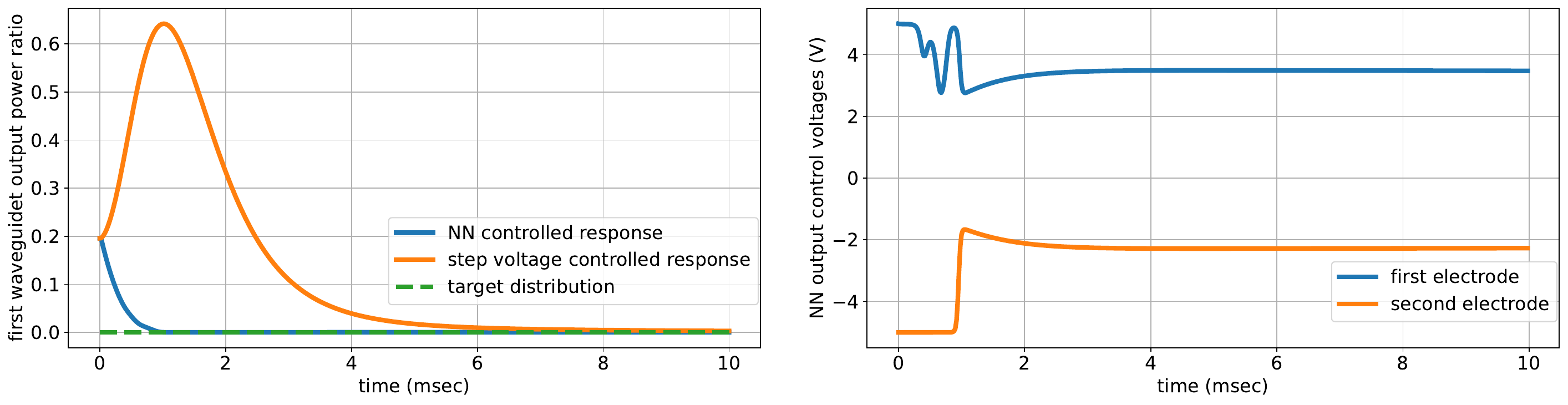}}
        \vspace{-1\baselineskip}
        \caption{target distribution $[0,1]$}
		\label{fig3:subfig1}
	\end{subfigure}
	\begin{subfigure}{1\linewidth}
		\centering{
	    \includegraphics[scale =0.348]{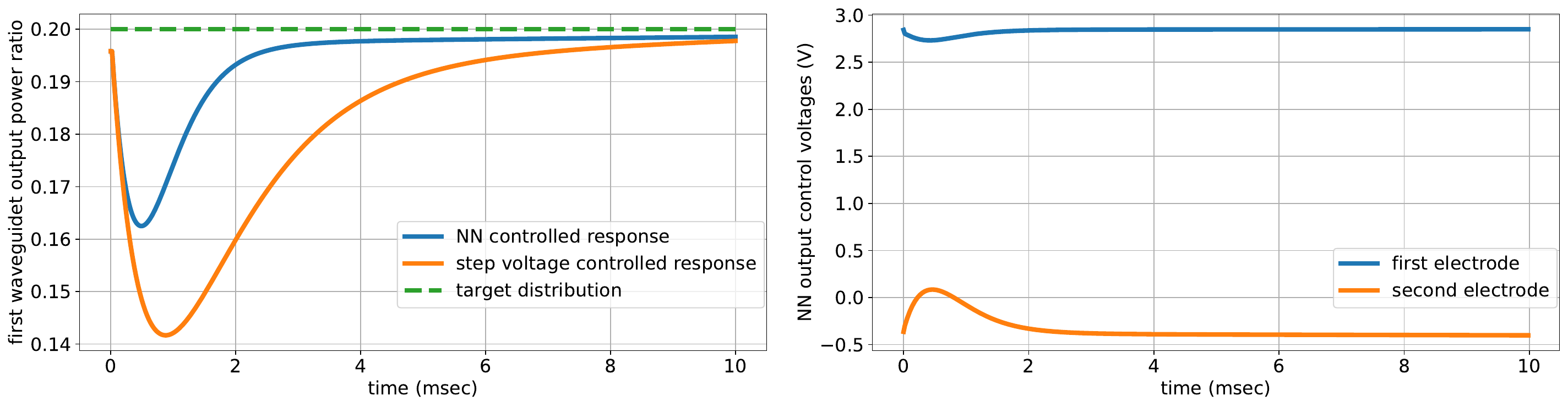}}
	    \vspace{-1\baselineskip}
	    \caption{target distribution $[0.2,0.8]$}
		\label{fig3:subfig2}
	\end{subfigure}
	\begin{subfigure}{1\linewidth}
		\centering{
	    \includegraphics[scale =0.348]{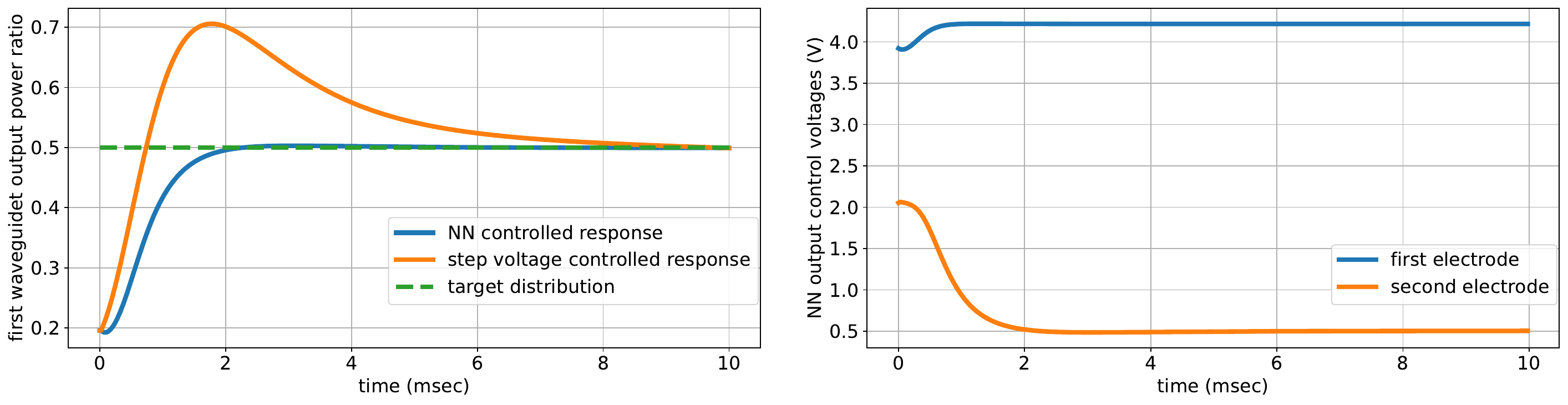}}
	    \vspace{-1\baselineskip}
	    \caption{target distribution $[0.5,0.5]$}
		\label{fig3:subfig3}
	\end{subfigure}
	\begin{subfigure}{1\linewidth}
		\centering{
	    \includegraphics[scale =0.348]{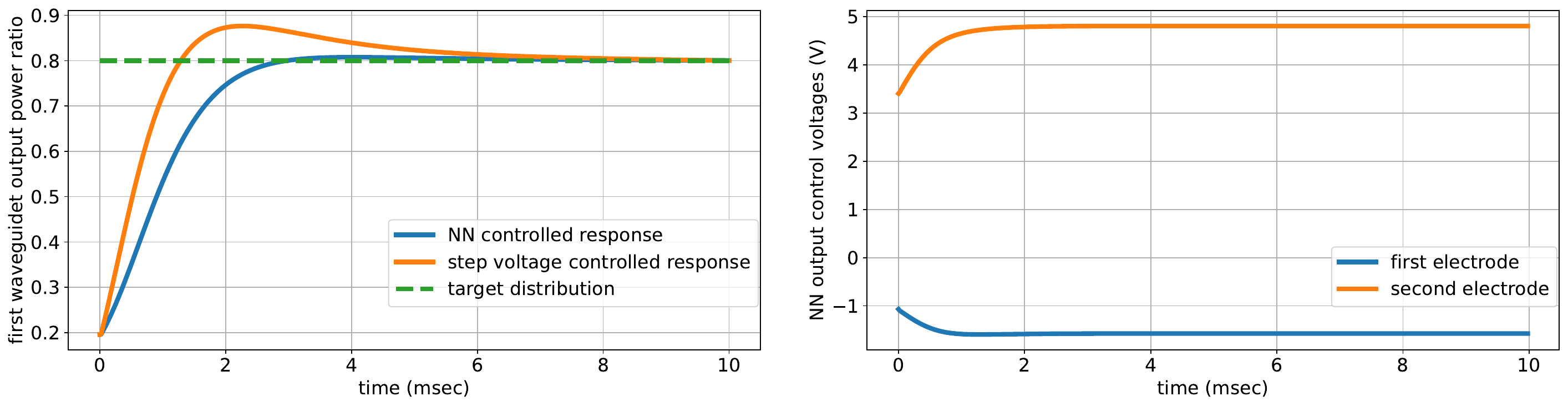}}
	    \vspace{-1\baselineskip}
	    \caption{target distribution $[0.8,0.2]$}
		\label{fig3:subfig4}
	\end{subfigure}
	\begin{subfigure}{1\linewidth}
		\centering{
	    \includegraphics[scale =0.348]{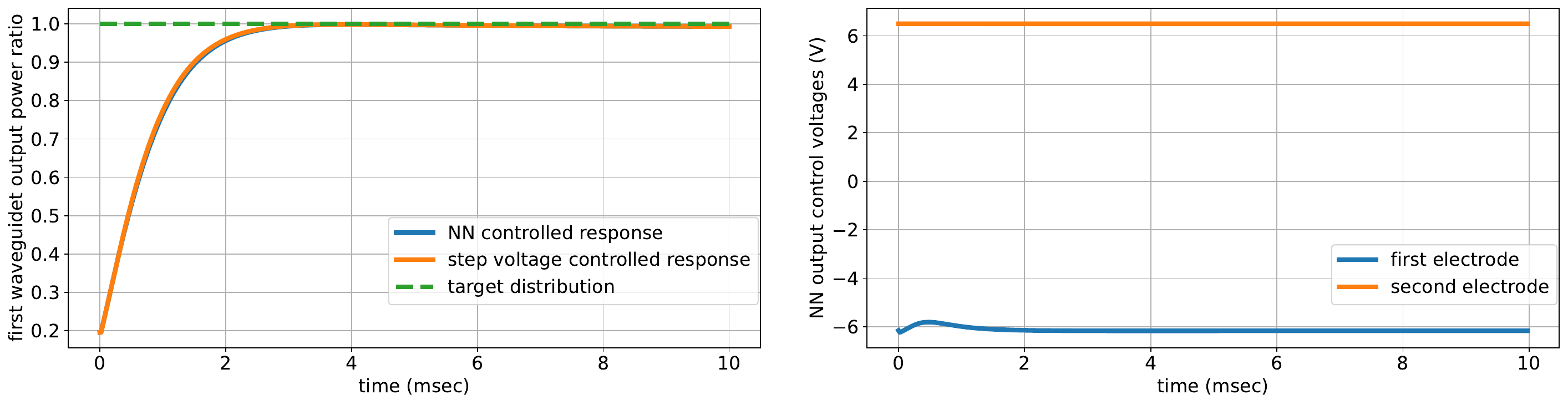}}
	    \vspace{-1\baselineskip}
	    \caption{target distribution $[1,0]$}
		\label{fig3:subfig5}
	\end{subfigure}
	\caption{The trained NN control the chip to bring its output to the corresponding target distribution within the episode time (10~msec) in comparison to applying a constant step voltages to the chip electrodes that could achieve the same target distribution within the same time limit. The left column is the first waveguide output power ratio. The right column is the control voltages generated by the trained NN and applied to the chip electrodes.
	}\label{fig3}
\end{figure}

We used an episode length $T$ of 500 steps, total episode time of 10~msec (i.e., each step is 0.02~msec), and sampling frequency of 2.5~MHz (i.e., the output power distribution of each step is sampled over 50 samples). We chose $N = 1$ and $\gamma = 0.99$. During the training of an NN, at the beginning of each episode, the chip (DRL enviroment) is reset to an initial state which is zero voltage being applied to the two electrodes, and the state $X$ of the LTI system representing the distortions is reset to [0,0]. We used Adam optimizer \cite{kingma2017adam} and L2 regularization \cite{regularization,Hastie_2020} with weight decay = 0.1. The weight initialization  scheme and learning rate $\eta$ used are different from one NN to another as shown in Table~\ref{tab1}. The learning rate was scheduled as the learning process advances. In the training stage, the action applied to the chip at each episode step is being randomly sampled from the gaussian distributions that are the output from the NN at the same step, while in the evaluation and operation stage, the action applied to the chip is the mean values of these gaussians, since they are the most probable suitable actions for the current input state to the NN. This allows more exploration for the DRL agent in the training stage. It should be noted that the output of the controller is limited to absolute value of 10~V during both training and operation. After finishing the training, we run an evaluation episode for each NN, where all the NNs were able to bring the chip output power distribution to the corresponding target distribution within the episode time (10~msec) with fidelity higher than 99\% as listed in Table~\ref{tab2}. The fidelity of the achieved output distribution is calculated as
\begin{equation}
	\text{fidelity} =((\sqrt{\alpha_{\text{achieved}}} \times \sqrt{\alpha_{\text{target}}}) + (\sqrt{\beta_{\text{achieved}}} \times \sqrt{\beta_{\text{target}}}))^2 \times 100\%\label{eq6}.
\end{equation}
Figure \ref{fig3} shows the performance of each trained NN in controlling the chip and bringing its output to the corresponding target distribution within 10~msec in comparison to applying constant step voltages to the chip electrodes that could achieve the same target output distribution within the same time limit. These constant step voltages were selected using grid search and are listed in Table \ref{tab3}.

\begin{figure}
	\begin{subfigure}[b]{0.33\linewidth}
		\centering
		\includegraphics[scale = 0.37]{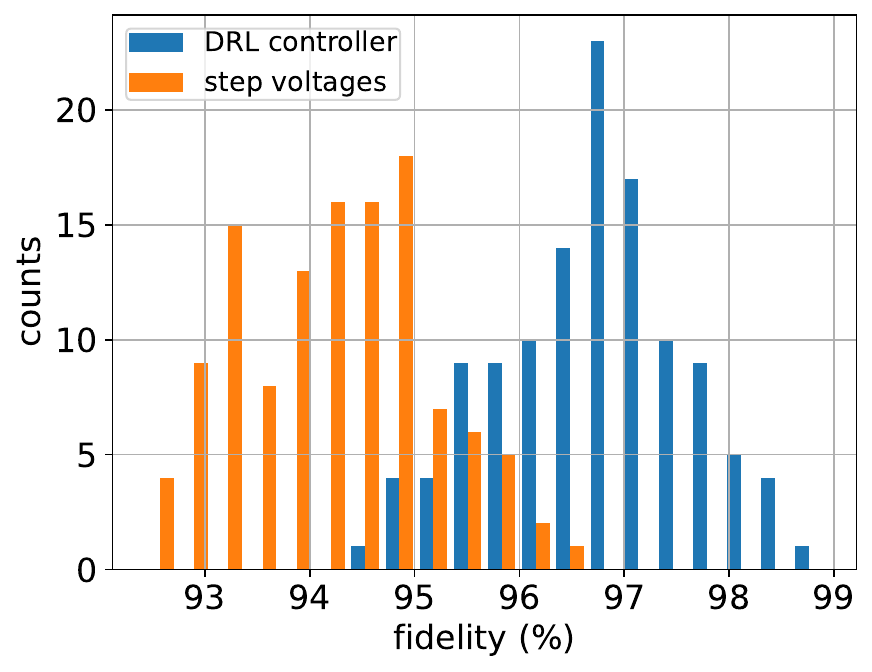}\caption{}\label{fig4:subfig1}
	\end{subfigure}
	\begin{subfigure}[b]{0.34\linewidth}
		\centering
		\includegraphics[scale = 0.37]{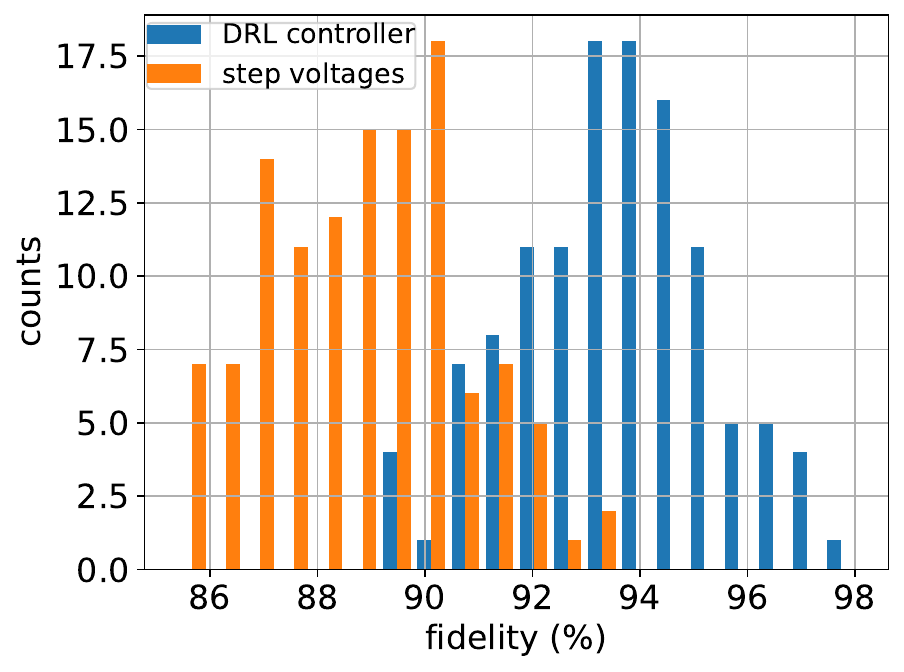}
		\caption{}
		\label{fig4:subfig2}
	\end{subfigure}
	\begin{subfigure}[b]{0.33\linewidth}
		\centering
		\includegraphics[scale = 0.37]{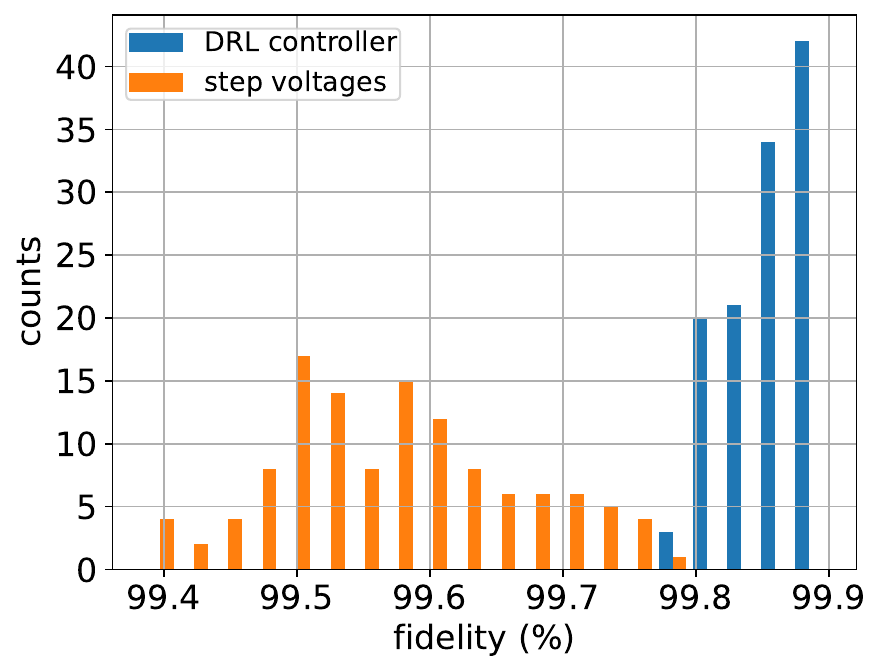}
		\caption{}\label{fig4:subfig3}
	\end{subfigure}
	\caption{Histogram for the fidelity of the sequences generated by the controller versus those generated by the step voltages listed in Table \ref{tab3}. The duration of each sequence is 50~msec (5 episodes), where each target distribution in the sequence lasts for 10~msec (1 episode). We generated all possible permutations of these target distributions which are 120 sequences. In (a) the fidelity is averaged over the whole sequence, in (b) the fidelity is averaged over the first 5~msec of each episode (which contain most of the transients) in the sequence, while in (c) the fidelity is averaged over the last 5~msec of each episode.}\label{fig4}
\end{figure}

\begin{figure}
	\centering
	\begin{subfigure}{1\linewidth}
		\centering
		\includegraphics[scale = 0.35]{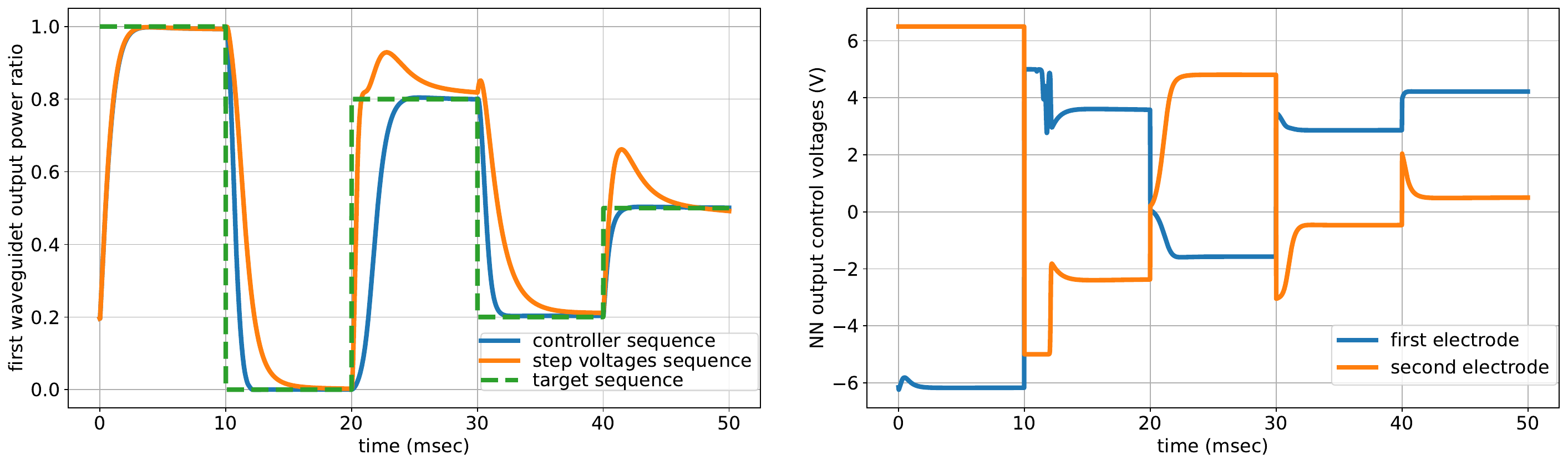}\caption{}
		\label{fig5:subfig1}
	\end{subfigure}
	\vfill
	\begin{subfigure}{1\linewidth}
		\centering
		\includegraphics[scale = 0.35]{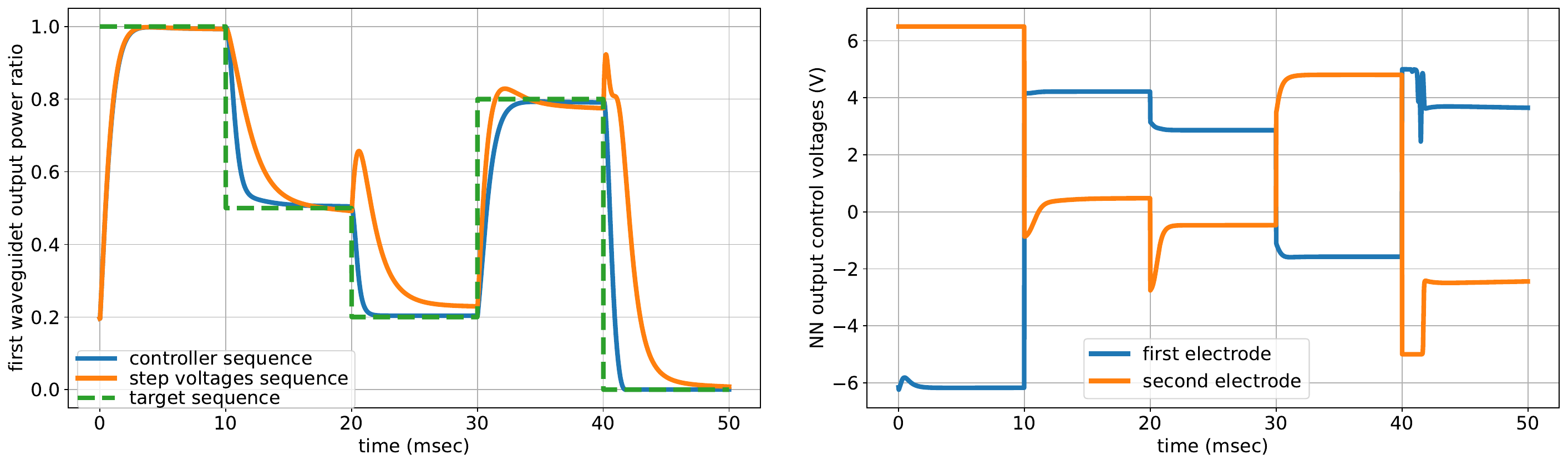}\caption{}
		\label{fig5:subfig2}
	\end{subfigure}
		\vfill
	\begin{subfigure}{1\linewidth}
		\centering
		\includegraphics[scale = 0.35]{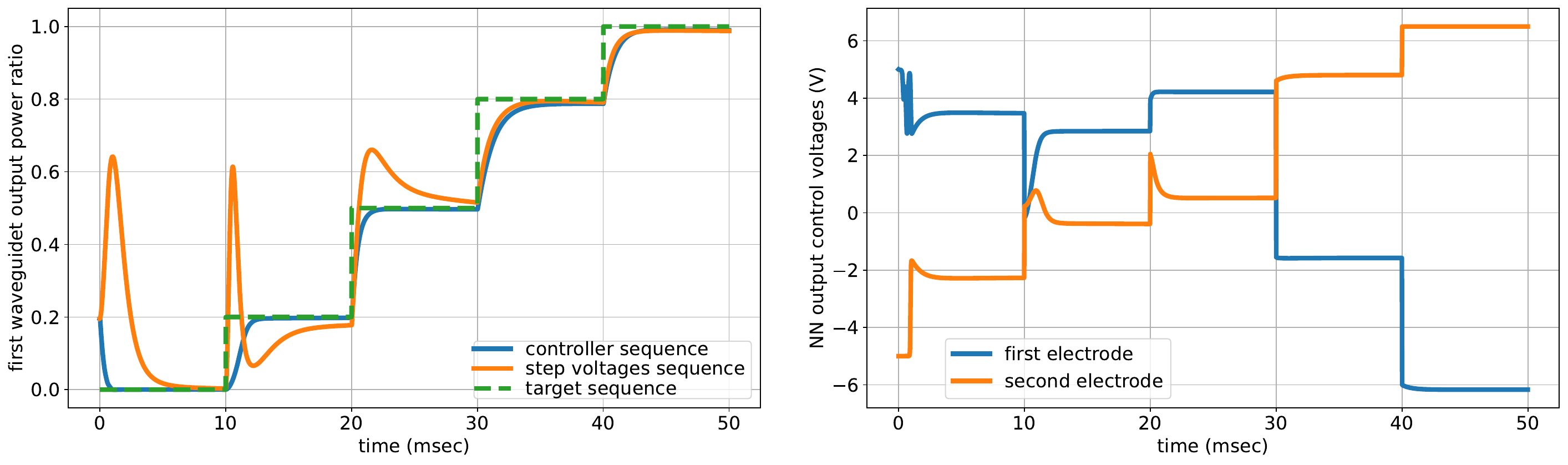}\caption{}
		\label{fig5:subfig3}
	\end{subfigure}
	\caption{Sequences generated by the controller versus the same sequences generated by the step voltages listed in Table \ref{tab3}. The duration of each sequence is 50~msec (5 episodes), where each target distribution in the sequence lasts for 10~msec (1 episode). (a) is the sequence with the lowest fidelity (averaged over the whole sequence), (b) is the one with the mean fidelity, while (c) is the one with the highest fidelity. The right column shows the first waveguide output power ratio. The left column shows the control voltages generated by the controller and applied to the chip electrodes.
	}\label{fig5}
\end{figure}

We conducted another experiment to assess the controller overall performance. We used our controller to control the chip to generate sequences of the five target output distributions we selected. The duration of each sequence is 50 msec (5 episodes), where each target distribution in the sequence lasts for 10~msec (1 episode). We do not reset the chip between episodes. We only reset the chip at the beginning of the sequence. We generated all possible permutations of these target distributions which are 120 sequences. Again we compared each sequence generated by the controller to the same one generated by the step voltages mentioned in Table \ref{tab3}. Figure \ref{fig4} shows a histogram for the fidelity of the sequences generated by the controller versus those generated by the step voltages. In Figure \ref{fig4:subfig1}, the fidelity is averaged over the whole sequence, in Figure \ref{fig4:subfig2}, the fidelity is averaged over the first 5~msec of each episode (which contain most of the transients) in the sequence, while in Figure \ref{fig4:subfig3}, the fidelity is averaged over the last 5~msec of each episode. The mean and standard deviation of each of the three cases are listed in Table \ref{tab4}. In Figure \ref{fig5}, we plotted the sequence with the lowest fidelity (averaged over the whole sequence), the one with the mean fidelity, and the one with maximum fidelity. We also plotted the corresponding control action generated by our controller in each case.
\begin{table}[h]
	\caption{Weights initialization scheme and learning rate used for each NN}\label{tab1}%
	\begin{tabular}{@{}llll@{}}
		\toprule
		NN & Target Distribution  & Initialization Scheme  & $\eta$\\
		\midrule
		$\text{NN}_{1}$    & [0,1]       & Xavier normalized initialization \cite{Xavier} with gain = 5                                & $7\times 10^{-5}$  \\
		$\text{NN}_{2}$    & [0.2,0.8]   & Kaiming normalized initialization \cite{kaiming} with hyperbolic tangent non-linearity       & $5\times 10^{-5}$  \\
		$\text{NN}_{3}$    & [0.5,0.5]   & Kaiming normalized initialization \cite{kaiming} with leaky relu non-linearity  			 & $8\times 10^{-5}$  \\
		$\text{NN}_{4}$    & [0.8,0.2]   & Xavier normalized initialization \cite{Xavier} with gain = 1.2 							 & $4\times 10^{-5}$ \\
		$\text{NN}_{5}$    & [1,0]       & Xavier normalized initialization \cite{Xavier} with gain = 2.2 							 & $5\times 10^{-5}$  \\
		\botrule
	\end{tabular}
\end{table}
\begin{table}[h]
	\caption{The evaluated fidelity achieved by each NN after training}\label{tab2}%
	\begin{tabular}{@{}llll@{}}
		\toprule
		NN & Target Distribution  & Fidelity \\
		\midrule
		$\text{NN}_{1}$    & [0,1]       & 99.99\%  \\
		$\text{NN}_{2}$    & [0.2,0.8]   & 99.99\%  \\
		$\text{NN}_{3}$    & [0.5,0.5]   & 99.99\%  \\
		$\text{NN}_{4}$    & [0.8,0.2]   & 99.99\% \\
		$\text{NN}_{5}$    & [1,0]       & 99.28\%  \\
		\botrule
	\end{tabular}
\end{table}

\begin{table}[h]
	\caption{Constant step voltages that could achieve the target distribution in 10~msec against which the the performance of the corresponding NN was evaluated}\label{tab3}%
	\begin{tabular}{@{}llll@{}}
		\toprule
		NN & Target Distribution  & Voltage [first electrode,second electrode] (V)  \\
		\midrule
		$\text{NN}_{1}$    & [0,1]       & $[-4.70,-2.90]$  \\
		$\text{NN}_{2}$    & [0.2,0.8]   & $[2.85,-0.40]$  \\
		$\text{NN}_{3}$    & [0.5,0.5]   & $[0.51,4.51]$  \\
		$\text{NN}_{4}$    & [0.8,0.2]   & $[-3.46,6.50]$  \\
		$\text{NN}_{5}$    & [1,0]       & $[-6.17,6.49]$  \\
		\botrule
	\end{tabular}
\end{table}

\begin{table}[h]
	\caption{The mean and standard deviation (controller vs. step voltages) of fidelity of the sequences generated by the controller versus those generated by the step voltages listed in Table \ref{tab3}}\label{tab4}%
	\begin{tabular}{@{}llll@{}}
		\toprule
		the period of each episode \\
		over which fidelity is averaged  & mean (\%)  & standard deviation(\%)  \\
		\midrule
		the whole episode    & 96.70 vs. 94.23       & 0.89 vs. 0.90  \\
		the first 5 msec     & 93.55 vs. 88.89       & 1.80 vs. 1.80  \\
		the last 5 msec      & 99.85 vs. 99.57       & 0.03 vs. 0.09  \\
		\botrule
	\end{tabular}
\end{table}

\section{Discussion}\label{sec5}
The presented results show the superior performance of our proposed controller in driving the waveguide array chip (introduced in \cite{Akramchip}) tackling the challenges mentioned in Section \ref{sec2}. Figure \ref{fig3} shows our controller excellent performance in controlling the chip transients (which is the most part affected by the classical signal distortions), where the controller makes the chip output settle faster at the target distribution if compared to just using step voltages, which showed significant overshoot and made the output took much longer time to settle. The superior performance of our controller in controlling the transients is evident as in Figure \ref{fig4}, where our controller exhibited a mean value of fidelity, averaged over all the generated sequences, of 96.7\% with a 3.2\% increase over the case of using step voltages for control. This difference even got bigger to be 4.7\% if we considered the first 5~msec only of each episode (which contains most of the transients) in the sequences. The mean value of fidelity in the histogram shown in Figure \ref{fig4:subfig1} for our controller is less than 99\%, since here the fidelity is averaged over the whole 10 msec of the episodes which include the transients part. However, if we considered the last 5~msec only of each episode in the sequence, the mean value of fidelity averaged over all the sequences will increase to 99.8\% (as shown in the histogram shown in Figure \ref{fig4:subfig3}), since the last 5~msec of each episode is in steady state (i.e., after the transient effect have diminished). This chip could be used for switching applications that frequenctly switch between target output distributions, and thus, reducing the transients effect is a requirement. During sequence generation, we do not reset the chip as we switch from a target distribution to another, which shows the ability of each NN in our controller to achieve its target distribution even if it started from a different state other than the reset state used in training as shown in Figure \ref{fig5}. This proves the ability of our controller to generalize quite well to situations unseen during training.
 
Through REINFORCE DRL algorithm, we were able to train a controller to control the chip without the need to model it at all. This controller was successful in generating voltage signals with proper value and waveform, as shown in the right column of Figure \ref{fig5}, to undo the signal distortions, which are the reason for the transients part in the output distribution of the chip, meanwhile achieving the intended target distribution with fidelity higher than 99\% in steady state. The control voltages were also limited to the desired operating range (from -10~V to 10~V).

The REINFORCE is a simple algorithm, easy to use, and straight forward to implement. This algorithm guarantees convergence to an optimal policy, which is a rare luxury in DRL algorithms, since it is a gradient ascent algorithm. This algorithm is an on-policy one, i.e., we need to generate new training samples using the most updated policy after each policy update (learning step). It also has an off-policy variant which is policy gradient with importance sampling (which also works for POMDP). This variant could be used if generating new data samples for each policy update during the training is not easy or monetarily expensive. However this variant is more computationally expensive due to importance sampling calculations. It also requires maintaining a replay buffer to store and sample past experiences, which increases the memory requirements.

The controller structure, we introduced, enabled covering as much target probability distributions as desired without increasing the complexity of training. We successfully trained five NNs for five different target distributions that spanned the chip output distribution from [0,1] to [1,0]. Extension to more distributions is just straightforward without introducing more difficulty or complexity to the control problem. During our experiments, we found that the neural network weight initialization scheme is crucial to have a successful training. We also tried Gated Recurrent Unit (GRU) NNs instead of the fully-connected NNs used in the controller. However, they were harder to train without having any extra benefits over the fully connected ones.

For this chip, we used continuous action space for the control signals, as they have continuous range of values. However, as we stated before, our method is suitable for discrete actions as well. In case of discrete actions, the NN will output the probability for each possible action in the discrete action space instead of outputting the mean and variance of a gaussian. 

The advantages our method has over the one used in \cite{Akramchip} are as follows. Our control method is a closed-loop one with feedback which enables the controller to compensate for disturbance or deviation affecting the system during operation. Our method is model-free, where the controller is trained directly on the system through direct interaction, and not trained on a pre-trained NN model of the system as in \cite{Akramchip}. This way we guarantee more accurate training, since in \cite{Akramchip} the NN model of the chip is trained to expect the chip output for specific voltage signals waveform (square pulses), while the controller NN is not constrained by any mean to generate square pulses. Moreover in our method, we do not need to design the training data set to guarantee generalization as in the deep learning scheme used in \cite{Akramchip}, because the training data samples are automatically generated during training by direct sampling from the actual environment.

\section{Conclusion}\label{sec6}
In this paper, we introduced a general method to control the state probability distribution of closed quantum systems as they evolve. We tackled two main common issues, which are the unmodeled classical distortions affecting the control signals, and the difficulty of modeling the quantum system itself. We used a model-free closed loop control scheme that applies REINFORCE policy gradient DRL algorithm, which works for both MDP and POMDP, to train a neural network as the controller. We proposed a novel architecture for the controller that can accommodate any number of desired target probability distributions, without increasing the complexity of the training process. Overcoming the issues mentioned earlier and with this controller architecture, our approach becomes suitable to handle most closed quantum systems. We validated our method on the quantum system introduced in \cite{Akramchip}, presenting the details of implementation in Section \ref{sec4}. The results showed high control performance of the proposed method. Since our control method is independent of the system dynamics, it can be directly extended to open quantum systems.

\bmhead{Acknowledgments}
This work was supported by the Australian Government through the Australian Research Council under the Centre of Excellence scheme (No: CE170100012).

\bibliography{sn-bibliography}

\end{document}